\documentclass[conference]{IEEEtran}
\usepackage{amsmath}%
\usepackage{amsfonts}%
\usepackage{amssymb}%
\usepackage{placeins}
\usepackage[pdftex]{graphicx}
\usepackage{nohyperref}
\usepackage{url}
\usepackage{fixltx2e}
\begin{document}

\title{Can network coding bridge the digital divide in the Pacific?}
\author{\IEEEauthorblockN{Ulrich Speidel, 'Etuate Cocker\\}
\IEEEauthorblockA{Dept.\ of Computer Science\\
The University of Auckland\\
\{ulrich$|$ecoc005\}@cs.auckland.ac.nz}
\and
\IEEEauthorblockN{P\'eter Vingelmann, Janus Heide\\}
\IEEEauthorblockA{Steinwurf ApS\\
Aalborg, Denmark\\
\{peter$|$janus\}@steinwurf.com}
\and
\IEEEauthorblockN{Muriel M{\'e}dard\\}
\IEEEauthorblockA{EECS \\
Massachusetts Institute of Technology\\
medard@mit.edu}
}
\maketitle
\IEEEoverridecommandlockouts
\IEEEaftertitletext{\vspace{-5cm}\begin{abstract}
Conventional TCP performance is significantly impaired under long latency and/or constrained bandwidth. While small Pacific Island states on satellite links experience this in the extreme, small populations and remoteness often rule out submarine fibre connections and their communities struggle to reap the benefits of the Internet. Network-coded TCP (TCP/NC) can increase goodput under high latency and packet loss, but has not been used to tunnel conventional TCP and UDP across satellite links before. We report on a feasibility study aimed at determining expected goodput gain across such TCP/NC tunnels into island targets on geostationary and medium earth orbit satellite links.
\end{abstract}}

\section{Introduction}
The present study has its genesis in a longitudinal project aimed at determining long-term trends in jitter, packet loss and other observables on long-distance Internet paths~\cite{iibex}. The project also documents the challenges faced by users in remote locations with numerous nodes in various Pacific Island nations. This naturally poses the question how one might improve their connectivity.

Bandwidth upgrades can address the bandwidth bottleneck to an extent, but do not address latency issues unless an island simultaneously upgrades to a lower latency connection, i.e., from a geostationary (GEO) to a medium earth orbit (MEO) satellite~\cite{O3b} or to a fibre-optic submarine cable.

Latency is particularly problematic when accompanied by packet loss. Satellite links to Pacific islands usually have much lower bandwidth than the international fibre networks supplying them, making them prime candidates for packet queue formation and hence for packet loss associated with tail drops.

Moreover, packet loss causes TCP sessions to lowering their data rate. However, long latency prevents TCP from doing so in a timely fashion, which can lead to queue oscillation at the satellite gateway~\cite{queueoscillation}. This can result in frequent idleness of the link, i.e., unused satellite bandwidth being observed even over short time intervals in the order of seconds.

Many small Pacific island states are developing nations with small populations (with similar numbers living overseas, mostly in New Zealand, Australia and the U.S.). Only a few, such as Hawaii, Fiji, and Guam, are the lucky transit points of international fibre routes. Others such as Tonga or Samoa are located close enough to such a transit point and populous enough to be able to afford a spur connection from a transit point. Many others simply do not have the population or budget for a fibre connection.

MEO satellites represent a lower cost/bandwidth alternative to fibre, with latencies roughly comparable to a fibre link from New Zealand to California. However, small islands often cannot muster the resources to install and maintain the required tracking antennas. At the time of writing, a small number of O3b MEO connections operate in Rarotonga, Samoa and East Timor~\cite{O3b}.

This leaves in particular the smallest, poorest, and most remote island communities reliant on expensive low-bandwidth and very long latency GEO satellite links. In many cases, thousands of people share an international link bandwidth equivalent to an average residential ADSL connection.

Various fast TCP variants such as H-TCP~\cite{htcp} and Hybla~\cite{hybla} attempt to establish a stable flow across long-latency connections, but are generally aimed at large-bandwidth scenarios to transfer large amounts of data efficiently. Our results indicate that they do not perform significantly better than conventional TCP in low-bandwidth scenarios in the Pacific and do not routinely outperform the most widely deployed Linux TCP variant, Cubic~\cite{cubic}.

Network coded TCP~\cite{nc-meets-tcp} (TCP/NC) is a forward error correction technique which may be used to hide packet losses between two hosts on a TCP network. This can mask tail drops at the satellite gateways, but an end-to-end TCP/NC solution requires both the sending and receiving host to ``speak'' TCP/NC. While this could conceivably be achieved for on-island hosts, it would be unrealistic to expect the rest of the Internet to follow suit.

An alternative solution is to deploy a network-coded ``tunnel'' between an off-island host with good connectivity to the Internet, and an on-island host acting as a gateway to an on-island network. This tunnel could mask satellite gateway losses and supply island end users with a low-loss connection to the Internet, avoiding the start-stop behaviour of their conventional TCP connections. But how well could this work in practice? This paper describes the initial results of a TCP/NC project in three Pacific Island locations with slightly different profiles.

We first give a brief birds-eye overview of TCP/NC and the basic network topology of our project, followed by an introduction to our deployment sites and initial observations from each.

\section{Network-coded TCP}
Network coding has attracted considerable research interest in information theory and communication engineering over the last decade, with much of the focus on solutions for mobile and wireless, especially in broadcast and multicast settings and in cooperative networking. In recent years, a number of authors have also proposed its use in conjunction with conventional point-to-point TCP/IP networking~\cite{nc-meets-tcp,hansenetal}. Network coding has also been investigated on the link layer of satellite links~\cite{bischl-et-al}. 

The fundamental idea of TCP/NC is to treat the original packets in a TCP connection as variables in a system of linear equations. As each packet is simply a binary string, it also represents a binary number, which can be multiplied and added. From a set of packets $p_1, p_2, \ldots, p_n$, {\em random linear network coding} (RLNC) generates a set of $n+\omega$ linear equations with random coefficients $c_{i,j}$ such that the $i$-th equation in the set is:
\begin{equation*}
\sum_{j=1}^{n}c_{i,j} p_j = r_i.
\end{equation*}
where $1\leq i \leq n+\omega$. We refer to $n$ as the {\em generation size} and to $\omega$ as the {\em overhead} and write $n+\omega$ to denote the size of the system. Instead of transmitting the original packets, TCP/NC transmits ``equation'' packets, i.e., the coded packet for the $i$-th equation encodes the $c_{i,j}$ for $1\leq j \leq n$ and $r_i$.

The receiving host can recover all of $p_1, p_2, \ldots, p_n$ once it has received $n$  linearly independent coded ``equation'' packets. Note that as the sender generates $n+\omega$ coded packets, the communication can in principle tolerate the loss of possibly up to $\omega$ arbitrary packets (readers without a background in linear algebra will probably still remember the ``at least $n$ equations for $n$ variables'' rule from school). Moreover, the sender can start generating redundancy as soon as two of the original packets are available, it does not have to wait for all $n$ packets like a block coder would have to.

The field from which one picks the $c_{i,j}$, the sparseness of the matrix $c_{i,j}$, the choice of $n$ and $\omega$ all influence robustness, coding rate, and decoding delay and complexity. A detailed discussion of these topics is beyond the scope of this paper. We merely note here that this is an active research area, some aspects of which we will return to in the rest of this paper.

\section{IP networking considerations for TCP/NC tunnels}
The tunnel in our experiments transports coded packets in the payload of UDP packets. UDP packets heading to the island originate from an off-island gateway $G_w$ and terminate at a gateway $G_i$ on the island. In this case, $G_w$ acts as the NC encoder and $G_i$ as the decoder. In the opposite direction, $G_i$ generates coded packets and $G_w$ is the decoder. Note that $G_w$ can be anywhere on the Internet: Rather than being topologically close to the satellite gateway $G_s$ facing the island, it should be close to the source/destination of the data required on-island to minimise the latency of the connection links covered by conventional TCP only.

\begin{figure}[htp]
    \resizebox{\columnwidth}{!}{\includegraphics{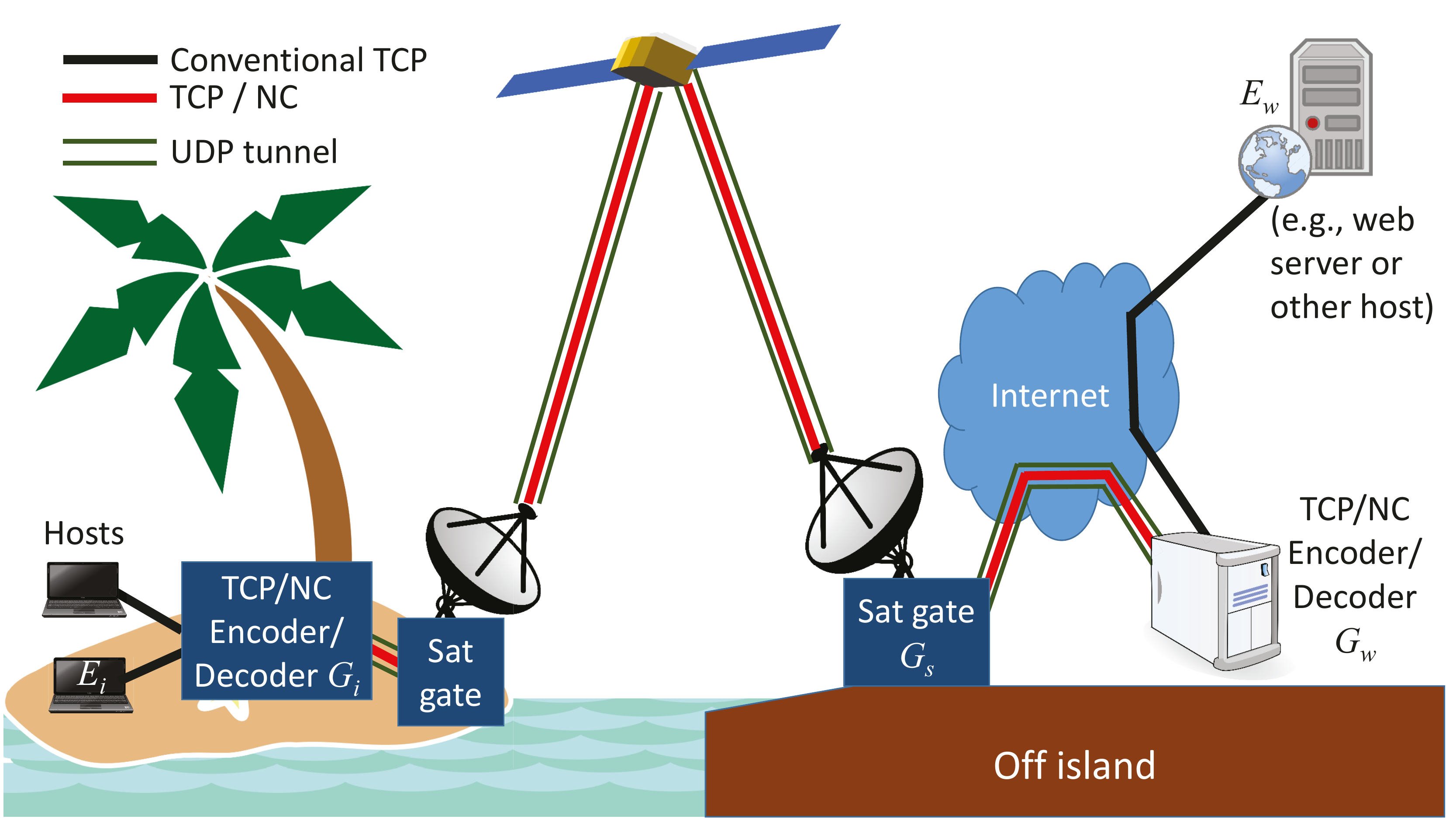}}
      \caption{TCP/NC network topology}
      \label{networktopology}
\end{figure}

The UDP headers of the tunnel packets contain only the IP addresses of $G_w$ and $G_i$. The IP addresses of the actual endpoints of each connection travelling through the tunnel travel inside the UDP payload as part of the NC header, i.e., they are not visible between $G_w$ and $G_i$. We denote these endpoints as $E_i$ (on-island) and $E_w$ (outside world).

For an IP packet from $E_i$ to reach $G_i$ for encoding, it suffices to insert $G_i$ as a ``gateway router'' for the on-island subnet in which $E_i$ resides. The tunnelled and decoded packet then egresses from $G_w$ enroute to $E_w$. N.B.: The network within which $G_w$ resides must allow egress of packets from $E_i$'s subnet.

In the opposite direction, an IP packet from $E_w$ to $E_i$ must find its way to $G_w$ for encoding. This requires one of the following two configurations to be implemented:
\begin{enumerate}
\item $E_i$ must be part of the autonomous system (AS) of $G_w$, and the AS must be configured internally to route traffic to $E_i$ via $G_w$ once it enters the AS.
\item Alternatively, $G_w$ may be one of the (at least) two BGP gateway routers giving access to an on-island AS within which $E_i$ resides. In this case, all gateways for this AS also need to be TCP/NC encoders/decoders to ensure that only encoded traffic enters the AS.
\end{enumerate}
Note that $E_i$ cannot simply use an address from the on-island network for which the satellite gateway $G_s$ acts as the IP gateway. If it did, BGP would route packets from $E_w$ to $E_i$ straight to $G_s$, bypassing the encoder $G_w$. In our experiments, we opted for the first configuration as it only needs a single off-island gateway $G_w$. The drawback of this solution is that the AS of $E_i$ starts off the island, potentially leaving an island ISP reliant on address blocks from the AS of a single offshore provider. Note also that the $G_i$'s IP address must be reachable via the satellite gateway $G_s$, so cannot be part of the same AS as $E_i$'s subnet.

Packet size also needs to be considered: As TCP/NC adds its own header to each encapsulated packet, the packet may exceed the MTU (maximum transmission unit) between $G_w$ and $G_i$, leading to IP fragmentation. While this does not affect principal functionality, it adds undesirable overhead and can affect timing and performance. For unfragmented UDP transit, the tunnel must limit the MTU for unencoded IP packets at $G_w$ and $G_i$ to the MTU of the path between $G_w$ and $G_i$ minus the size of the NC header.

In our experiment, we use a $G_w$ with two Ethernet interfaces, {\em eth0} and {\em eth1}, configured as follows:
\begin{itemize}
\item {\bf eth0} is the tunnel interface. Its single exclusive route points at the ``normal'' island IP address of $G_i$'s tunnel interface only so it can reach it via $G_s$. {\em eth0} connects to its own router port in its own exclusive /30 subnet. The tunnel endpoint itself is a virtual network device supplied by Steinwurf ApS as a loadable Linux kernel module. This endpoint intercepts packets from $G_i$'s tunnel interface arriving at {\em eth0} and decodes them. Similarly, any unencoded packets sent to the virtual adapter are encoded and sent to $G_i$ via {\em eth0}.
\item {\bf eth1} is the default ``world'' interface. It also connects to its own router port in its own exclusive /30 subnet, separate from {\em eth0}'s subnet. Any outgoing traffic not destined for {\em eth0}'s subnet or $G_i$ egresses here. Similarly, the outside world routes traffic destined for $G_w$ and for $E_i$'s on-island subnet to {\em eth1}. Any IP packets destined for the latter subnet are sent to the virtual network device for encoding.
\end{itemize}
The $G_w$ in our study is a Dell PowerEdge R320 server with an Intel Xeon E5-2420 v2 Processor running at 2.20~GHz based at the University of Auckland. A further $G_w$ endpoint is planned for California.

At the island ends, the respective $G_i$ have three Ethernet interfaces (say {\em eth0}, {\em eth1} and {\em eth2}):
\begin{itemize}
\item As in $G_w$, {\em eth0} is the tunnel interface with a single exclusive route pointing at the IP of {\em eth0} on $G_w$. The tunnel endpoint is the same kernel module as in $G_w$, configured to peer with {\em eth0} on $G_w$ to intercept UDP from there for decoding. Any unencoded packets sent to the virtual adapter are encoded and sent to $G_w$ via {\em eth0}.
\item {\em eth1} now serves as the interface to $E_i$'s on-island subnet. Decoded traffic for $E_i$ egresses here, and any traffic arriving at this interface from the outside is sent to the kernel module for encoding.
\item {\em eth2} acts as a ``maintenance interface'' for emergency remote access from off-island machines via $G_s$ if {\em eth0} is unavailable. {\em eth2} and {\em eth0} must be in different networks. Policy routes let $G_i$ respond to incoming standard TCP traffic on both {\em eth0} and {\em eth2}.
\end{itemize}
Our $G_i$ at each island end is a Stealth LPC-630F Little PC industrial computer with an Intel i7-3520M Gen 3 processor.

\begin{figure*}[tph]
    \resizebox{\textwidth}{!}{\includegraphics{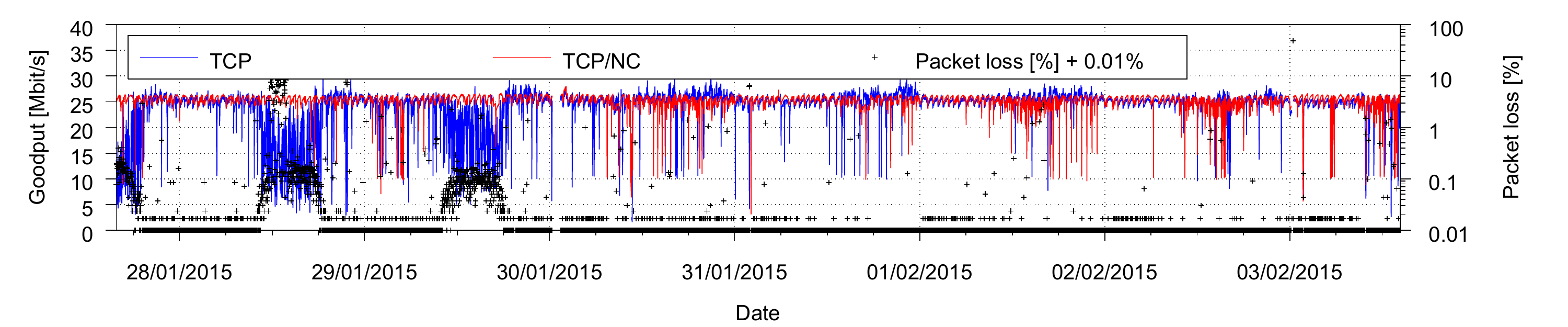}}
        \caption{Comparison between TCP and TCP/NC goodput over time using a $30+6$ TCP/NC tunnel between Auckland and Rarotonga. Note the comparatively stable performance of TCP/NC at times of low TCP goodput.}
        \label{raro20pc}
\end{figure*}

\section{Deployment sites}
The study deployed $G_i$ in the following locations:
\vspace{2mm}

\noindent
{\bf Rarotonga, Cook Islands}. Rarotonga has a permanent population of over 13,000, with an estimated 1000 to 3000 visitors at any time~\footnote{Authors' estimate based on monthly government visitor arrival statistics~\cite{ckislandsarrivalstats}.} Telecom Cook Islands (TCI), Rarotonga's sole Internet provider, connects primarily via the O3b MEO satellite network. At time of deployment, TCI subscribed to 160 Mbps down- and 40 Mbps uplink, normally routed via O3b's Hawaii teleport. Casual 802.11b/g/n Internet retailed for around US\$ 0.06 - 0.08 per MB on Rarotonga; US\$ 75 bought a month of ADSL connection limited to 4 Mbps with an 8 GB data cap.
\vspace{2mm}

\noindent
{\bf Niue}. This single-island country has a permanent population of around 1,600 with only a small number of visitors but a large overseas diaspora: Over 20,000 Niueans live in New Zealand alone~\cite{nupopstats}. Niue's only ISP is Internet Niue, with a GEO connection now providing 8 Mbps down- and 2 Mbps uplink after an upgrade. Niue has a free local public access 802.11 WiFi network covering most of the island. US\$ 70 buys a month of ADSL limited to 384 kbps / 2 GB.
\vspace{2mm}

\noindent
{\bf Funafuti Atoll, Tuvalu}. Funafuti is the most populous atoll in Tuvalu with $\approx$4,500 inhabitants. The sole ISP, Tuvalu Telecommunication Corporation (TTC), uses a GEO link. Its satellite provider also manages some of TTC's on-island network near the local satellite gateway for TTC. The GEO downlink rate into Funafuti was not available to us, but our measurements indicate a rate of 16 Mbps. US\$100 buys 256~kbps ADSL with 3 GB data/month.

A somewhat unusual component of TTC's network is a SilverPeak NX-3700 WAN Optimiser~\cite{silverpeak}, which carries all production IP traffic. Details on the configuration of the device were not available. However, its functions include forward error correction in tunnels via parity packets that can be added automatically or at pre-set ratios to protect against sporadic packet loss. We also observed that the device had network memory activated. Other features include packet reordering, coalescing of smaller packets into larger ones, IP header / payload compression, and TCP and other protocol acceleration.

Network memory in particular makes it challenging to compare TCP/NC and conventional TCP across a SilverPeak path, as some packets are served from local cache and never transit the link. We therefore positioned the TCP/NC encoder / decoder in Funafuti such that our traffic to it did not pass through the SilverPeak device. However, we measured conventional TCP goodput to a machine supplied through the NX-3700.
\vspace{2mm}

Both Niue and Tuvalu have signed up to the Kacific geostationary high-throughput broadband satellite system~\cite{kacific}, with a ready-for-service date in 2017. This will increase the bandwidth into both countries to levels broadly comparable to the current Rarotonga connection and will provide direct-to-building services.

%\FloatBarrier
\section{Observations to date}
Data from our IIBEX database, collected using VoIP-like UDP traffic with small packets transmitted at 50 packets/s, showed packet loss on both up- and downlink to and from Rarotonga of between 0 and 0.25\% before deployment. To and from Niue, packet loss was below 0.1\%. No recent data was available for Tuvalu, but data from 2013 suggested packet losses of up to 2.5\%. At deployment time, we encountered similar values with significant variation over time.

\begin{figure}[htp]
    \resizebox{\columnwidth}{!}{\includegraphics{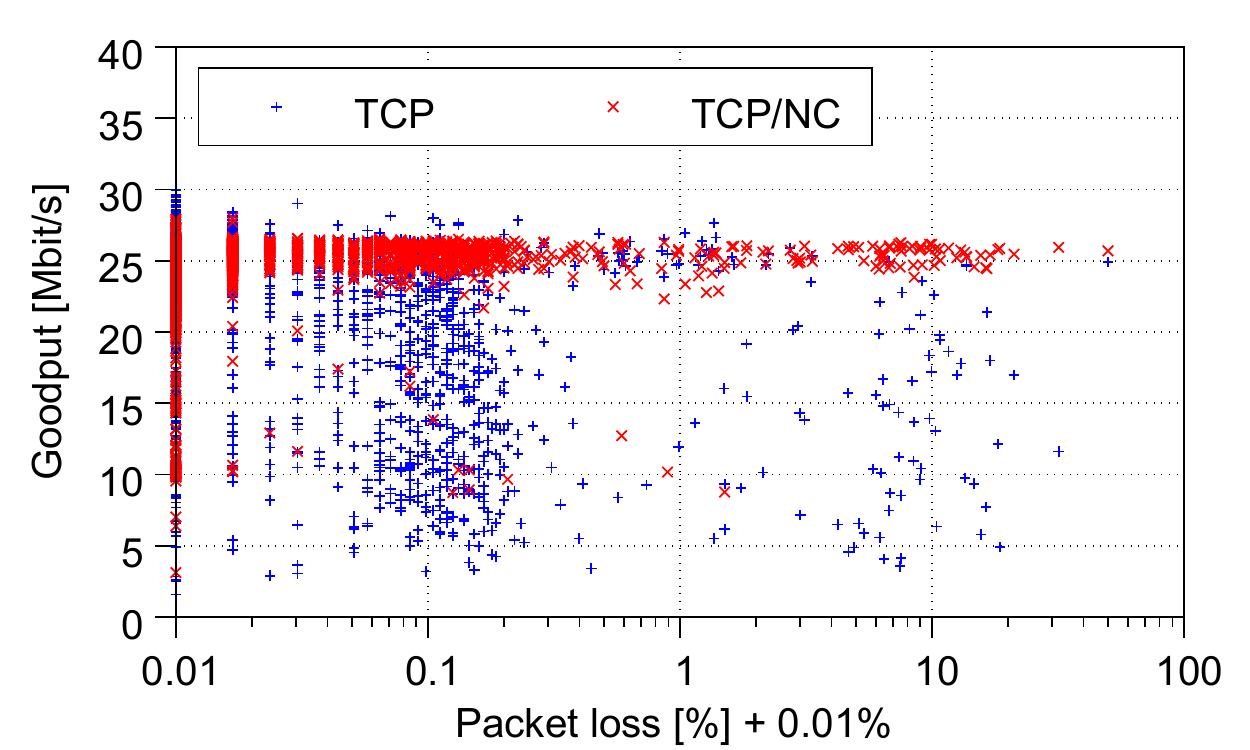}}
      \caption{TCP and TCP/NC goodput vs.\ packet loss using a $30+6$ TCP/NC tunnel between Auckland and Rarotonga. TCP/NC almost always outperforms TCP, which always does worse when packet loss is high.}
      \label{raro20pcscatter}
\end{figure}

As part of the deployment, we undertook extensive throughput measurements in all locations using both conventional TCP and TCP/NC. We found the following in the respective locations:
\vspace{1mm}

\noindent
{\bf Rarotonga}: Conventional TCP seemed unable to access more than around 75\% of the available satellite bandwidth, even over short time intervals in the order of seconds. Most of the time, the observed utilisation was below 60\%. As satellite uplinks are not subject to medium contention, the utilisation below 100\% implies that the input queue to the satellite modem in Hawaii {\em empties} completely during those time periods.

However, we also observed regular burst packet losses in inbound traffic. These do not occur between the various offshore locations involved, indicating an association with the satellite link. Burst losses are characteristic for tail drops on {\em overflowing} queues, in this case most likely the input queue to the satellite modem, where high bandwidth fibre or Ethernet traffic feeds into the lower bandwidth of the satellite link, but can also be caused by the radio link itself.

Link underutilisation with this behaviour is a well-known effect called {\em queue oscillation}~\cite{queueoscillation}, which often occurs at Internet bottlenecks with multiple parallel flows. All satellite links under consideration here qualify as bottlenecks with bandwidths below those of the networks connected at either end. Queue oscillation can occur when multiple TCP flows try to ``fill the pipe'' through the bottleneck. Traffic banks up until the queue overflows, resulting in packet loss. The TCP senders cannot detect this until an ACK becomes overdue or a selective repeat request arrives. Neither of these events reduces the queue arrival rate for a full end-to-end round-trip-time (RTT) period, long enough to turn most connections' packet losses into bursts. This causes all senders to slow down and the arrival rate at the queue now drops substantially below the satellite bandwidth; the queue drains and the link sits idle. Without further packet losses, the sending rates increase again. This cycle can repeat within only a few RTTs, explaining why the effect is seen even at small timescales of just a few seconds.

Initially, we required $n+\omega=60+30$ (=50\% redundancy) to mask the packet loss bursts for our TCP sender to sustain a higher packet rate. Between Ubuntu servers offshore and a Linux client on the island's test network, goodput improved by up to a factor of 4 with the TCP/NC tunnel. Bandwidth utilisation increased to almost 90\% for a single download via the tunnel. With a Windows 7 client, goodput improvement was only $\approx 25\%$, which was sustainable for several parallel connections, however. Possible reasons may include fragmentation or packet reordering across the tunnel.

When revisiting the Rarotonga link in late January 2015, we found that (a) the total downlink bandwidth had increased to 200 Mbps, (b) conventional TCP now slightly outperformed the $60+30$ tunnel, (c) the queue still seemed to be oscillating, with similar utilisation percentages, i.e., around 25\% additional throughput, and (d) a $30+6$ tunnel (constant 20\% overhead) was now sufficient to mask losses in most cases and gave better goodput than TCP, especially at times of high packet loss (see Fig.~\ref{raro20pc}). During periods of low or no packet loss, the $30+6$ tunnel mostly yielded goodput comparable to conventional TCP. This suggests that an adaptive overhead scheme could replace most of the overhead by goodput during these times.
\vspace{1mm}

\noindent
{\bf Niue}: The link into Niue sees sustained peak data rates of around 7.5~Mbps with $>$7~Mbps recorded for much of the day. Individual conventional TCP connections achieve around 0.3~Mbps. Packet loss into Niue is low but not zero. Closer inspection reveals that the link transports goodput without redundant retransmissions arriving at the Niue end. At the utilisation and data rates observed, the link can thus handle around 25 parallel connections.

Single TCP connections across a TCP/NC tunnel achieved around 2-2.4~Mbps goodput with very low overhead, i.e., the entire link capacity would be exhausted by 3-4 such TCP/NC connections. Given the high existing link utilisation, this additional performance of even a single connection comes at the expense of conventional TCP goodput. However, if these connections are downloads, the higher goodput rate also shortens the flow. This poses the question as to whether short wideband flows with TCP/NC are better from a user perspective than long thin ones, given that the bulk of bandwidth use consists of flows that download something.

In Niue, we also investigated the potential of H-TCP and Hybla compared to the standard Cubic TCP used by Ubuntu. While there were considerable differences between them and Cubic at certain times, neither of the two presented a convincingly strong alternative on this narrowband path.
\vspace{1mm}

\noindent
{\bf Funafuti}: No link utilisation data was available, but we were able to measure traffic directly between the sat gate and TTC's local network with a line tap and nprobe/ntop~\cite{ntop}. Without TCP/NC, utilisation for total IP data traffic into Funafuti seldom exceeded 2-3 Mbps -- less than 20\% of available bandwidth. Packet losses occurred as soon as a relatively modest load was offered. In combination, these observations once again suggest queue oscillation. 

Conventional TCP to the Funafuti TCP/NC encoder/decoder (i.e., not through the NX-3700) did not exceed 1.6~Mbps at any time. During peak hours, rates were $<$0.4 Mbps and some connections timed out. Closer analysis revealed that peak time packet loss bursts during oscillation ran into the hundreds of packets -- longer than the maximum overhead presently possible in our TCP/NC kernel module. Downloads into Funafuti via TCP/NC tunnel achieved a steady average goodput of around 4 Mbps for a single connection. Link utilisation with a $60+30$ generation tunnel observed regularly reached short-term peak rates of over 15~Mbps, i.e., longer downloads may well reach average rates $>$4~Mbps. Conventional TCP throughput on the local network side of the NX-3700 reaches up to 13~Mbps. Given the comparatively low data rate observed as going into Funafuti, it is likely that a significant part of the 13~Mbps is supplied by the NX-3700 network memory function. 

\section{Conclusions and future work}
This paper reported on preliminary observations from our three deployment sites. These show that TCP/NC can provide significantly higher goodput rates for individual connections than conventional TCP and that it can exploit spare capacity on satellite links left idle due to queue oscillation. On such links, TCP/NC can provide additional goodput. On links already carrying mostly goodput (Niue), it can merely replace long and more or less parallel data transfers by short and mostly successive ones. A possible benefit of this is that individual downloads complete faster, reducing user temptation to abort downloads (and hence waste partially downloaded data).

There are many obvious questions that we have not yet been able to investigate and many challenges to address. These include finding optimal generation sizes and overheads, adapting them to conditions, investigating the potential use of a performance-enhancing proxy (PEP) that breaks up the TCP connection into multiple sections, and the performance of TCP/NC if used in conjunction with network memory and other functions of a WAN optimiser. Also, we still expend a lot of TCP/NC overhead on a problem primarily caused by the conventional TCP traffic our TCP/NC streams mingle with. Would an all-TCP/NC link work with less overhead? That said, the pace at which these investigations will be able to proceed is limited by the very problem we are trying to solve: the long-latency, low-bandwidth satellite links into Pacific Islands.

\section*{Acknowledgements}
This research was supported by the Information Society Innovation Fund Asia through the Pacific Island Chapter of the Internet Society (PICISOC) and by Internet New Zealand. We would also like to thank the many Internet users and staff of Telecom Cook Islands, Internet Niue, and the Tuvalu Telecommunication Corporation for their patience during this study and for sharing their precious bandwidth with us.

\end{document}